\documentclass[12pt,preprint]{aastex}
\usepackage{ragged2e}
\def\beq{\begin{equation}}
\def\eeq{\end{equation}}
\def\beqar{\begin{eqnarray}}
\def\eeqar{\end{eqnarray}}

\def\la{\mathrel{\mathpalette\fun <}}

\def\fun#1#2{\lower3.6pt\vbox{\baselineskip0pt\lineskip.9pt
  \ialign{$\mathsurround=0pt#1\hfil##\hfil$\crcr#2\crcr\sim\crcr}}}

\def\mrm{\mathrm}

\begin{document}

\title{DIFFUSE PIONIC GAMMA-RAY EMISSION FROM LARGE-SCALE STRUCTURES IN THE \emph{FERMI} ERA}

\author{ A. Dobard\v zi\'c}
\affil{Department of Astronomy, Faculty of Mathematics, University of Belgrade, Studentski trg 16, 11000 Belgrade, Serbia}
\email{aleksandra@matf.bg.ac.rs}

\and

\author{T. Prodanovi\'c}
\affil{Department of Physics, University of Novi Sad, Trg Dositeja Obradovi\'ca 4, 21000 Novi Sad, Serbia}
\email{prodanvc@df.uns.ac.rs}

\begin{abstract}
For more than a decade now the complete origin of the diffuse gamma-ray emission background (EGRB)
has been unknown. Major components like unresolved star-forming galaxies (making $\la 50\%$ of the EGRB) and
blazars ($\la 23\%$), have failed to explain the entire background observed by \emph{Fermi}. Another, though
subdominant, contribution is expected to come from the process of large-scale structure formation. The growth
of structures is accompanied by accretion and merger shocks, which would, with at least some magnetic field
present, give rise to a population of structure-formation cosmic rays (SFCRs). Though expected,
this cosmic-ray population is still hypothetical and only very weak limits have been placed to their
contribution to the EGRB. The most promising insight into SFCRs was expected to come from \emph{Fermi}-LAT
observations of clusters of galaxies, however, only upper limits and no detection have been placed. Here, we build a model
 of gamma-ray emission from large-scale accretion shocks implementing a source evolution
calibrated with the \emph{Fermi}-LAT cluster observation limits.
Though our limits to the SFCR gamma-ray emission are weak (above the observed EGRB) in some case, in others, some
of our models can provide a good fit to the observed EGRB.
 More importantly, we show that these large-scale shocks could still give an important
 contribution to the EGRB, especially at high energies. Future detections of cluster gamma-ray emission would help place tighter constraints on our models
and give us a better insight into large-scale shocks forming around them.
\end{abstract}


\section{INTRODUCTION}
\label{sec:intro}
The extragalactic gamma-ray background (EGRB) was first detected by the {\it OSO-3} satellite \citep{oso72}, confirmed by the {\it SAS-2}
satellite \citep{FS78, TF82}, which was later
succeeded by the {\it EGRET} \citep{ST00}, and has most recently been measured by the  \emph{Fermi}-LAT
\citep{ABD10} to unprecedented precision. In the search for the origin of the EGRB, multiple classes of sources
were considered. Unresolved normal galaxies were found to be dominant contributors to the EGRB
\citep{FPP10,LHB12}. Other sources like blazars \citep{SS96,NT07,DE07,IT09,SPA12}, pulsars \citep{HL76,HS81,FL10}, dark matter annihilations \citep{SC10,ABDO10}, or secondary gamma-ray cascades \citep{MBT12,II12} were also considered. Even before \emph{Fermi} era, \cite{PA93} considered gamma-ray emission from cores of radio galaxies as a contributor to the EGRB for energies $>100\,\mathrm{MeV}$ based on \emph{Compton Gamma Ray Observatory} data. More recently, the contribution of radio galaxies using \emph{Fermi} observations were considered in \cite{IN11}. In most cases, none of these nor their combinations, managed to explain the entire
background, though in a few models, some parameters can result in gamma-ray emission above the observed values.
Analysis of the anisotropies in the EGRB \citep{AC12} and cross-correlations with source catalogs \citep{XIA11} can be
used to constrain and discriminate between contributions of these unresolved sources.
In terms of the guaranteed sources \citep{PF02}, \emph{Fermi} observations revealed that the dominant emission mechanism is
through cosmic-ray
interaction with the gas in the interstellar medium (ISM), which produces neutral pions $\pi^0$ which then decay into gamma rays
$\mathrm{p}_\mathrm{cr}+\mathrm{p}_\mathrm{ism}\to\pi^0\rightarrow\gamma+\gamma$ \citep{STE70,STE71}. This process,
where cosmic rays are accelerated in supernova remnants, is the dominant process responsible for the gamma-ray emission of
observed star-forming galaxies \citep{ST76,PF02,PF06,MTK11} and contributes to the EGRB via unresolved star-forming
galaxies. An accompanying leptonic component, namely inverse-Compton emission, was also analyzed, however it
turned out to be subdominant to the hadronic component in star-forming galaxies contributing to the EGRB
\citep{CF12}. Though normal galaxies are the most natural source of pionic gamma-ray emission because of their
ongoing star-formation process that is constantly injecting cosmic rays into the ISM, quiescent galaxies rich
in gas where cosmic rays are injected from Type Ia supernova events were also analyzed and were found to be
potentially important contributors to the EGRB in some cases \citep{LF12}.

However, as cosmic rays are accelerated anywhere where shocks and magnetic fields are present,
other cosmic-ray populations accelerated at different sites might also contribute to the observed EGRB. One
such population are the cosmic rays accelerated during large-scale structure formation (SFCRs; Loeb \& Waxman 2000;
Furlanetto \& Loeb 2004; Miniati et al. 2000), however, they are still hypothetical and yet to be observed. Good
sites for observing potential gamma-ray emission from SFCRs are the nearby galaxy clusters
\citep{VO96,BE97,EN97,CB98,KW09,PP10}, however observations have only placed upper limits so far by \emph{Fermi}
\citep{AC10}, and thus SFCRs remain undetected and their contribution to the EGRB remains unknown, with
only weak limits placed \citep{MI03,PF04,PF05,KU05}. The search for emission
from nearby galaxy clusters was also performed at $\mathrm{TeV}$ energies by telescopes like MAGIC \citep{ALEK10}, \emph{H.E.S.S.} \citep{AH09,AHA09}, CANGAROO-III \citep{KI09}, HEGRA \citep{AAB06}, and Whipple telescope \citep{PBB06}, but again only upper limits have been reported. SFCRs should also be detectable with radio observations, since some merging galaxy clusters were already detected in radio waves \citep{BR11,VW11,MM11}. Although cosmic rays that arise from processes responsible for growth of structures are yet to be detected, using in concert observations and limits at different energies will be important in revealing their signatures.

Nonetheless, given the importance of disentangling the origin of the EGRB and identifying its dominant components,
in this work we build an improved and more detailed model of SFCR contribution to the EGRB.
Since the history of this cosmic-ray population is unknown, as its tracer, we implement a semi-analytical source evolution of
accretion shocks \citep{PF06}. Another unknown required to make a prediction of their collective emission
contributing to the EGRB is the gamma-ray flux expected from a single, average, large-scale shock. Though the
cosmological shocks vary in scale and type (merger, accretion, filament), in our model, we will assume that
accretion shocks around galaxy clusters are the dominant type of large-scale shocks, which allows us then to use cluster gamma-ray
detection limits placed by \emph{Fermi}.  A similar approach was used in \citet{KU05} to obtain the contribution of galactic clusters
to the EGRB observed by \emph{EGRET} but with a simpler source evolution.
We note that only the pionic gamma-ray emission from SFCRs will be analyzed since it has been shown to be
dominant over the inverse-Compton emission \citep{CF12,PP10}.

\section{FORMALISM}
\label{sec:form}

The differential gamma-ray intensity $dI_E / d\Omega$  $({\rm cm^{-2} s^{-1} GeV^{-1} sr^{-1}})$
is an observable quantity that describes the EGRB. Following Pavlidou \& Fields (2002, hereafter
PF02) and \cite{PF04} the differential gamma-ray intensity coming from SFCRs is
\begin{equation}
\label{eq:1}
\frac{dI_E}{d\Omega}=\frac{c}{{4\pi H_0}}
\int{\frac{{\dot{n}_{\gamma,\mathrm{com}}[z,(1+z)E]}}{{\sqrt{\Omega_\Lambda +\Omega_\mathrm{m}(1+z)^3}}}dz}\,,
\end{equation}
where $H_0$ is the present value of the Hubble parameter and $z$ is the redshift of the source. Matter and
vacuum energy density parameters are given as $\;\Omega_\mrm{m}$  and $\Omega_{\Lambda}$, respectively.
The differential co-moving gamma-ray emissivity density is $\dot{n}_{\gamma,\mathrm{com}}$. In the case of
gamma-ray emission from normal galaxies, where emission comes from cosmic rays accelerated in supernova
remnants,  $\dot{n}_{\gamma,\mathrm{com}}$ is a function of the number density of galaxies and their individual
emission. This can be related to cosmic star-formation rate $\dot{\rho}_*(z)$
as $\dot{n}_{\gamma,\mathrm{com}} = L_{\gamma} n_\mathrm{gal} \propto \dot{\rho}_*(z)$ (see Equation (4) of PF02). In the case of SFCRs where
the sources are large-scale structure formation shocks, their gamma-ray emissivity density,
$\dot{n}_{\gamma,\mathrm{com}}$, can be expressed in terms of a similar quantity which we call the cosmic
accretion rate, $\dot{\rho}_\mrm{sf}(z)=\int d{\cal M} d\dot{\rho}_\mrm{sf}(z,{\cal M})/d{\cal M}$
(co-moving mass current density crossing the surfaces of shocks of all Mach
numbers at a given cosmic epoch in units of $M_\odot \rm yr^{-1} Mpc^{-3}$). Analytical models of cosmic accretion shocks were
constructed by Pavlidou \& Fields (2006, hereafter PF06)
using the double distribution formalism of \cite{PAFI05}.
In their work, PF06 analyze cosmic shocks that arise on different accretors (with Press--Schechter
mass distribution), at different epochs, in environments of different local over- and underdensities.
They calculate the power and mass current that
enters into these shocks of various strengths and follow their evolution over redshift, taking into account
the effect of the environment such as the preheating. Furthermore, PF06 demonstrate that their models are consistent with
energetics of accreted matter that results from simulations. We will use the model of PF06 to implement the
evolution of SFCR sources, for which we thus implicitly take only the cosmic accretion shocks.
The gamma-ray emission resulting from a cosmic-ray population scales as a product
of cosmic-ray flux and total mass of targets $L_{\gamma} \propto \phi_{\rm cr} M_{\rm gas}$.
We note, however, that the underlying assumption for which this relation holds is that the gas and cosmic-ray
distributions are homogeneous within a system. Though this is true for galactic diffuse gamma-ray emission,
for the case of large structures such as galaxy cluster, where we expect to find this cosmic-ray population, this would
be most valid for the peripheral regions \citep{keshet2010,Mur08}. More precise modeling of expected gamma-ray emission of a given
accretor would have to include a thermal gas profile of the intra-cluster medium \citep{PFEN04}, and a more valid assumption would
be that cosmic rays trace thermal gas distribution, which would alter the above relation and include additional parameters.
However, for the purpose of this work, where cluster gamma-ray emission upper limits will be utilized to constrain the
accretion shock contribution to the EGRB, the details of a specific emission of a single accretor are not important, but will be included
in a more extensive follow up analysis.

In the case of the galactic cosmic rays, their flux can be taken to be proportional to the
star-formation rate, while similarly, in the case of cosmic rays originating from accretion shocks, the cosmic-ray flux can
be taken to be proportional to the mass accretion rate (mass current crossing the shock surface in units $M_{\odot} {\rm yr^{-1}}$)
 of shocks at some object and some epoch, $J(z)$.
With that set, a gamma-ray luminosity from structure formation process at some redshift can be determined as
\begin{equation}
\label{eq:2}
L_\gamma(z,E)=\frac{J(z)}{J_0(z_0)}\frac{M_{\rm gas}(z)}{M_{\rm gas}(z_0)}L_{\gamma,0}(E)\,,
\end{equation}
where $E$ is photon energy in the accretor rest frame and $M_{\rm gas}(z)$ is gas mass contained in the accretor at
a given cosmic epoch $z$, i.e., the mass of the intracluster gas. $J_0$ is the accretion rate at $z_0$ to which we normalize. The gamma-ray luminosity of the normalization
cluster, taken as the cosmic average, is then $L_{\gamma,\mathrm{0}}$. The implicit assumption in the above equation is that the
ratio of accelerated to accreted particles  is a constant. The emissivity density can thus be written as
$\dot{n}_{\gamma,\mathrm{com}}(z,E)=L_{\gamma}n_\mrm{c}$. Co-moving galaxy cluster number density, $n_\mrm{c}$,
and cosmic accretion rate, $\dot{\rho}_\mrm{sf}(z)$, are connected via $\dot{\rho}_\mrm{sf}(z)=J(z) n_\mrm{c}$.
This gives the relation
\begin{equation}
\label{eq:3}
\dot{n}_{\gamma,\mathrm{com}}(z,E)=L_{\gamma,\mathrm{0}}[(1+z)E]\frac{\dot{\rho}_\mrm{sf}(z)}{J_0(z_0)}
\frac{M_{\rm gas}(z)}{M_{\rm gas}(z_0)}\,.
\end{equation}
Assuming that, on the onset of accretion when structure was virialized, there was some initial gas mass $M_{\rm gas,0}$ defined
with respect to the accreted gas as  $M_{\rm gas,0}=\epsilon M_{\rm gas,acc}(z_0)$, the above gas mass ratio can be
written in terms of the accreted mass ratio as
\begin{equation}
\label{eq:4}
\frac{M_{\rm gas}(z)}{M_{\rm gas}(z_0)}=\frac{\epsilon+ M_{\rm gas,acc}(z)/M_{\rm gas,acc}(z_0)}{1+\epsilon},
\end{equation}
where $M_{\rm gas,acc}(z)$ and $M_{\rm gas,acc}(z_0)$ are masses of gas accreted from the epoch of
virialization up to the redshifts $z$ and $z_0$, respectively. The ratio of the accreted masses is equal to the
ratio of the cosmic accretion rates during those same epochs
\begin{equation}
\label{eq:5}
\frac{M_{\rm gas,acc}(z)}{M_{\rm gas,acc}(z_0)} =
\frac{\int_{z_\mathrm{vir}}^{z} dz\left(dt/ dz \right)
\dot{\rho}_\mrm{sf}(z)}{\int_{z_\mathrm{vir}}^{z_0} dz\left(dt/ dz\right)
\dot{\rho}_\mrm{sf}(z)}\,.
\end{equation}

Finally, following PF02, Equations (\ref{eq:1}), (\ref{eq:2}), (\ref{eq:3}), (\ref{eq:4}) and (\ref{eq:5}) combine to give
the SFCR gamma-ray intensity as
\begin{eqnarray}
\label{eq:main}
\frac{d I_E}{d\Omega}&=& \frac{c}{4\pi H_0 J_0(z_0)} \int_0^{z_\mathrm{vir}}
dz\frac{\dot{\rho}_\mrm{sf}(z)L_{\gamma,0}[(1+z)E]}{\sqrt{\Omega_\Lambda +\Omega_\mathrm{m}(1+z)^3}}\, \nonumber\\
&\times& \left[ \frac{\epsilon}{\epsilon+1}+ (\epsilon+1)^{-1}\frac{\int_{z_\mathrm{vir}}^{z}dz\left(dt /dz \right)
\dot{\rho}_\mrm{sf}(z)}{\int_{z_\mathrm{vir}}^{z_0}dz\left(dt/ dz\right)\dot{\rho}_\mrm{sf}(z)} \right]\,,
\end{eqnarray}
where our solutions depend on the initial gas fraction parameter $\epsilon$, and the assumed spectral index, $\alpha_\gamma$.

\section{INPUT}
\label{sec:input}

\subsection{Cosmic Accretion Rate}
\label{sec:accrate}

For the total mass current density of gas entering into accretion shocks at a given epoch,
which we call the cosmic accretion rate $\dot{\rho}_\mrm{sf}$, we use models constructed in PF06. These
models present an analytical description of the energetics of the population of cosmic accretion shocks.
We utilize their derived mass current distribution among different shock Mach numbers and their evolution
with cosmic time, as a tracer of structure formation shock history, and thus a tracer of SFCR history.
We note however, that the results of PF06 that we use as cosmic accretion rate, $\dot{\rho}_\mrm{sf}$, reflect
the evolution of a distribution of shock strength at each redshift, while in our work we assume that most
emission comes from typical objects representative of the cosmic mean, with a single shock strength at each redshift.

Three models are separately analyzed in PF06. The simplest assumes that all objects are embedded in an
environment well represented by the background universe. This model is based on the Press--Schechter formalism and will be
labeled as Model 1 in our results.
The second model of PF06 includes variations in the local matter density and temperature in the region
around the accretor that are imprinted in the primordial density field and will be labeled as Model 2. The third, most realistic model, also includes
filament preheating and compression. This model will be labeled as Model 3 in our results. The latter two models that include
environmental effects are based on a
double distribution which describes how the number density of collapsed and virialized dark matter objects
is distributed among different masses and among different local density contrasts with respect to the
cosmic mean density \citep{PAFI05}.

Since galactic cosmic rays are accelerated in supernova remnants, and star-formation rate reflects on the supernova rate,
 the cosmic star formation rate, which reflects the evolution of the star formation rate, can then be taken
to reflect the evolution of galactic cosmic-ray sources.
In a similar manner, if we assume that SFCRs are accelerated in accretion shocks, and that the integrated (over all accretion
shocks of all Mach numbers) mass current density derived in PF06 models reflects the evolution of accretion
shocks, then we can take that integrated mass current density reflects the evolution of SFCR source, i.e., take it as the cosmic
accretion rate, $\dot{\rho}_\mrm{sf}$.

\subsection{Gamma-Ray Spectra}
\label{sec:gspec}

For the shape of the pionic gamma-ray spectrum $\Gamma_{\gamma , \pi^0}(E)$, we used the semi-analytical formula
derived by \cite{PFEN03} as a representation of Dermer's model \citep{DE86}. The
spectrum in logarithmic space is symmetrical around half the pion rest mass, $m_{\pi^0}$, with the slope
of the spectrum at high-energy end reflecting the spectral index $\alpha_{\gamma}$ of cosmic rays, which
we will leave as a free parameter in our model.

Even though large hopes were placed on \emph{Fermi}-LAT when it comes to detection of galaxy clusters in
gamma rays \citep{PP10}, no detections have been made yet. During the 18 months of \emph{Fermi}-LAT observations, 33 clusters
of galaxies were investigated and only upper limits have been reported \citep{AC10}. Still, one can use the
reported upper limits to set the normalization of the SFCR component of the EGRB, which itself
is then the upper limit.
We choose to normalize to the Coma cluster as a typical cluster and thus use
its \emph{Fermi}-LAT upper limit, as well as other corresponding parameters entering Equation
 (\ref{eq:main}) such as the Coma redshift
$z_0=0.0232$ \citep{CH07} and $z_\mathrm{vir}=1.5$ \citep{B11} for its virialization. To determine $L_{\gamma,0}$ we
again assume a \cite{PFEN03} gamma-ray spectral shape $\Gamma_{\gamma ,
\pi^0}(E)$, the virial
mass of Coma cluster $M_{500}=9.95\times10^{14}\,M_{\bigodot}$ (Chen et al. 2007; assumed to be
the total mass of Coma cluster $M_\mrm{tot}$), and the virial radius
$r_{500}=1.86\,\mrm{Mpc}$ \citep{CH07} which is defined as the radius at which the interior mass density equals
$500\rho_\mrm{c}$ , where $\rho_\mrm{c}$ is the critical density at the redshift of the cluster. For the gas
mass $M_\mrm{gas}$ of Coma cluster, we use the value $M_{\mrm{gas},500}=19\times
10^{13}\,M_{\bigodot}$ \citep{CH07}.
With that set, and adopting the \emph{Fermi} detection limit
of Coma as actual detection with gamma-ray flux of
$F_{\gamma,0}=4.58\times10^{-9}\mrm{photon\,cm^{-2}\,s^{-1}}$ \citep{AC10} integrated over the energy range
$0.2 - 100\,\mrm{GeV}$, the normalization of the luminosity spectrum of Coma can be derived by
requiring that  $F_{\gamma,0} = \int dE\,L_{\gamma,0}(E) / 4(1+z)d_c(z)^2\pi = C \int
dE\,\Gamma_{\gamma , \pi^0}(E) / 4(1+z)d_c(z)^2\pi$ where again $\Gamma_{\gamma , \pi^0}(E)$ is the shape of
pionic spectrum, $C$ is the spectrum normalization constant and $d_c(z)=97\,\mathrm{Mpc}$ is the co-moving distance of the cluster
standardly defined as $d_c(z)=(c/H_0) \int_0^z dz'/ \sqrt{\Omega_M(1+z)^3+\Omega_\Lambda}$.
Hence, the cluster luminosity is found to be $L_{\gamma,0}(E)=5.85\times10^{47}\,
\Gamma_{\gamma , \pi^0}(E)\,\mrm{photon}\;\mrm{s}^{-1}\,\mrm{GeV}^{-1}$.

Finally, one has to worry about how attenuation of gamma-ray photons by the extragalactic background light (EBL; Salamon \& Stecker 1998; Stecker et al.
2006; Kneiske et al. 2004; Gilmore et al. 2009, 2012; Razzaque et al. 2009; Finke et al.
2010; Abramowski et al. 2013) will affect our results.
Namely, high-energy gamma rays get attenuated by interacting with the EBL which results in the electron--positron pair production.
EBL consists of photons with wavelengths
from ultraviolet to infrared, that come directly from stars or have been absorbed and reprocessed by dust in their host galaxies.
\emph{Fermi}
observations of blazars and gamma-ray bursts have resulted in upper limits for EBL \citep{ABD09,ABDOO10} and have detected an
EBL redshift-dependent signature \citep{ACK12}. EBL attenuation was also recently measured in the spectra of the brightest blazars observed by  \emph{H.E.S.S.} \citep{ABR13}.
Since the electron--positron pair production cross section peaks at twice the electron mass, gamma-ray photons of energies $\sim\mathrm{GeV}$ are most likely to interact with UV background photons. Thus, models with high UV backgrounds will result in more suppression at high energies.
The attenuation is redshift and energy-dependent and is significant at energies $E>100\,\mathrm{GeV}$.
Although gamma-ray attenuation at higher redshifts is still highly uncertain,
using the fiducial model for gamma-ray attenuation from \cite{GSP12}, we find that the modeled SFCR emission starts
dropping after about $E=100\,\mathrm{GeV}$. This trend becomes even bigger after $1000\,\mathrm{GeV}$, and after
$10,000\,\mathrm{GeV}$, no photons can reach us. This means that the attenuation of SFCR-produced gamma-ray photons
by EBL might be visible when higher energy observations from \emph{Fermi} become available.

One should also bare in mind that electron--positron pairs produced as a result of attenuation of gamma rays,
could then inverse Compton scatter off the cosmic microwave background radiation and result in a secondary gamma-ray
(cascade) emission component \citep{AH94,FDW04}.  This process results in a redistribution of gamma-ray photons from the high to low-energy end.
Therefore, the cascade emission is especially important for
hard gamma-ray sources whose emission extends to high energies. This type of emission is expected to contribute to the EGRB
\citep{CA97,IT09,II12,MBT12} and should even start to dominate over the EBL-attenuated primary gamma-ray
component at $\mathrm{GeV}-\mathrm{TeV}$ energies.
It was shown that the cascade emission can also depend on the strength of the intergalactic magnetic field \citep{YZZ12,VP13}.
Namely, as the cascade particles get deflected in the magnetic field, this results in a ``smearing out" the point source emission and could,
with sufficiently strong magnetic fields, result in decreased brightness of a source to the point where it becomes unresolved, and its gamma-ray emission
gets ``merged" with the background.
Since the intergalactic magnetic field is
still unknown and only upper and lower limits exist \citep{NS09}, and since the cascade emission also depends on the spectrum of primary gamma rays which is a free parameter in our model, we at this point omit the calculation of cascade emission and present only primary gamma-ray emission produced by SFCR with attenuation included.

\section{RESULTS}
\label{sec:res}

The contribution of SFCR interaction to EGRB was derived from Equation (\ref{eq:main})
based on the semi-analytical model of evolution of accretion shocks and the \emph{Fermi}-LAT detection limit of
the Coma cluster. The cosmological parameters used were $\Omega_\Lambda=0.7$, $\Omega_\mrm{m}=0.3$, and
$H_0=71\;\mrm{km\,s^{-1}\,Mpc^{-1}}$.
For our default case and results plotted in Figure \ref{fig:2} we adopt initial gas mass parameter,
$\epsilon=0$, and a spectral index typical of strong shocks, $\alpha_\gamma=2.1$
(we find the mean Mach numbers weighted by the accreted gas current for all three environment models of PF06 to all be
consistent with strong shocks).

Our results are plotted in Figure \ref{fig:2} where we plot contributions of different components to the EGRB flux (data points)
detected by \emph{Fermi} \citep{ABD10}. The upper panel shows each component separately, while the bottom panel sums over
components and gives the total predicted EGRB.
The dash dotted line (red curve in the online version of the article) and dash dot dotted line (blue curve in the online version of the article)
represent the contribution from star-forming galaxies in the two limiting cases of the model --
luminosity evolution and density evolution respectively \citep{FPP10}.  The luminosity evolution scenario takes the redshift evolution of the
galaxies in such a way so that luminosity of the galaxies was allowed to evolve while their co-moving number density was kept constant.
Pure density evolution is the case
where evolution lies in co-moving number density of normal galaxies while their luminosity is constant.
The solid line is the blazar contribution \citep{AAA10}. Contribution from SFCRs based on three
different source models of PF06, and calibrated with Coma cluster detection limit,
is represented by a long dashed (Model 1), short dashed (Model 2) and dotted curve (Model 3). The same line types relate to the same
models on the bottom panel as well. The bottom panel shows the summed contribution from all components--blazars, star-forming
galaxies, and SFCRs. Thick curves (and colored red in the online version of the article) correspond to the luminosity, while thin curves
(and colored blue in on-line version of the article) correspond to the density evolution limiting
case of the star-forming galaxy contribution to the EGRB as given in \cite{FPP10}.
The EBL attenuation is also included, which impacts our results at $E>100\,\mathrm{GeV}$.
\begin{figure}\centering
{\label{fig:2a}
\includegraphics[width=0.65\textwidth]{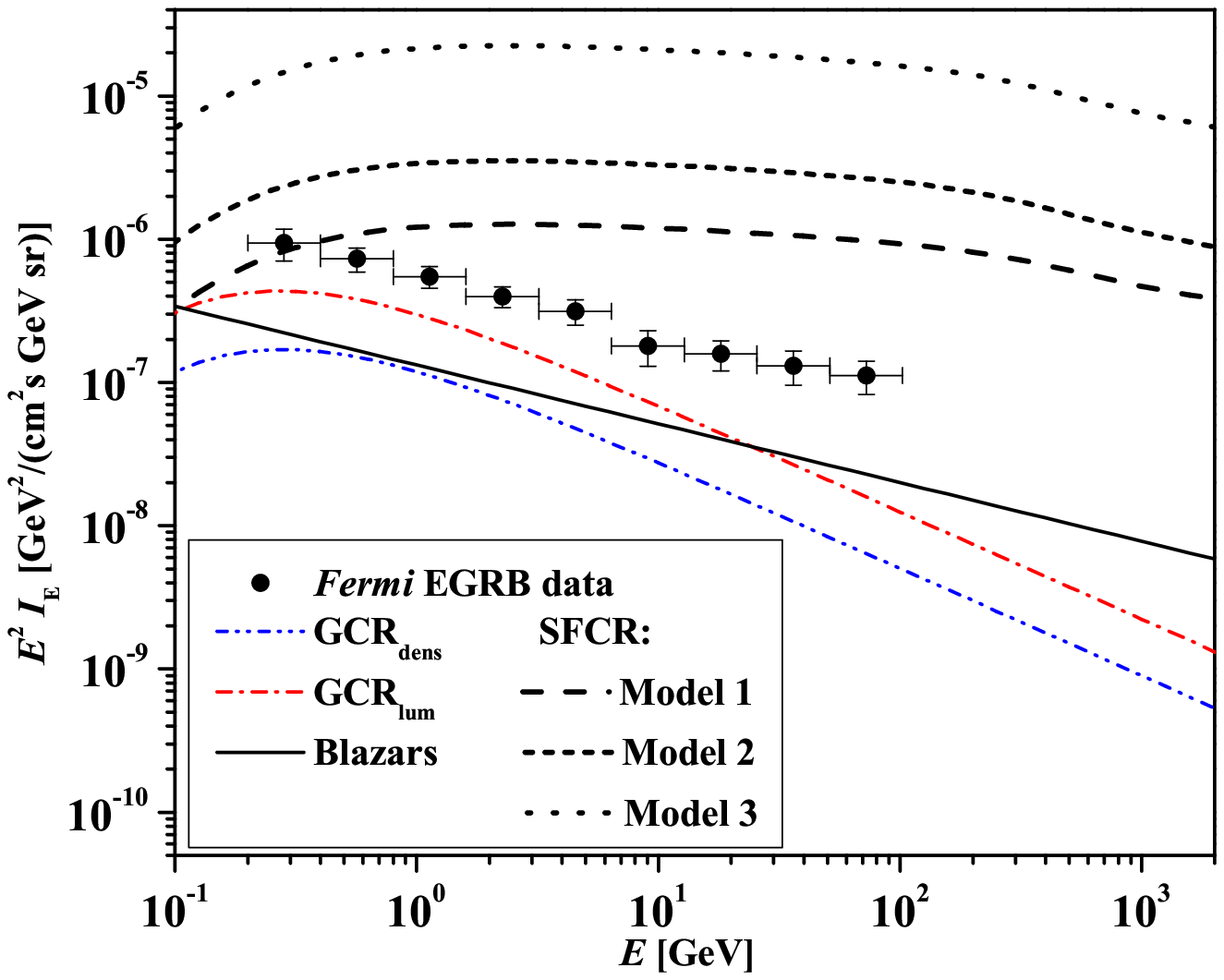}}
{\label{fig:2b}
\includegraphics[width=0.65\textwidth]{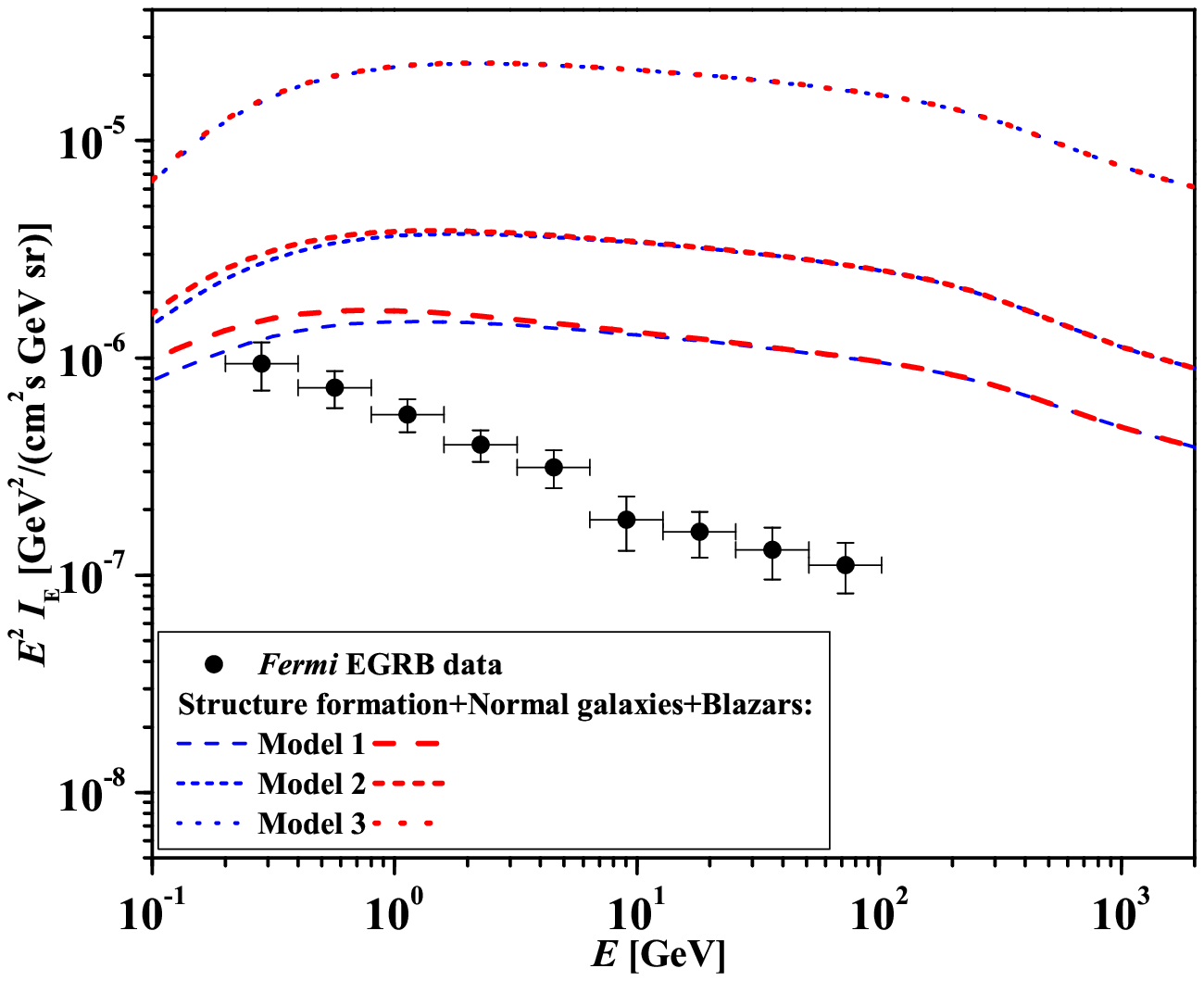}}
\centering
\caption{\label{fig:2}
Contribution of different components to the
EGRB (data points) observed by \emph{Fermi} \citep{ABD10}.
Top panel: all components shown separately -- blazars (solid line), normal star-forming galaxies based on two limiting cases given in
Fields et al. (2010; red dash dotted line represents luminosity evolution and blue dash dot dotted line represents density evolution), and structure-formation cosmic-ray
contribution calculated in this work,
normalized to the Coma cluster gamma-ray flux limit, with spectral index assumed to be $\alpha_\gamma=2.1$, and
initial gas mass parameter $\epsilon=0$, for three different source models derived in Pavlidou \& Fields (2006; long dashed, Model 1; short dashed, Model 2; dotted line, Model 3).
Bottom panel: the combined contribution of all components where different curves reflect different
normal galaxy emission models (thick red curves, luminosity evolution; thin blue curves, density evolution)
and different structure-formation cosmic-ray emission models
(three different line types correspond to the same models as on the top panel).
\newline
(A color version of this figure is available in the online journal.)}
\end{figure}
What is apparent is that estimated diffuse gamma-radiation from SFCRs differs by more than one order of magnitude depending on
which environment
model was used; Model 1, with no environmental effects where collapsed structures accrete gas of uniform density and
temperature, results in the lowest
gamma-ray emission, while Model 3, where environmental effects were included (filament preheating) and virialized structures are located within
preheated gas, results in the highest gamma radiation. This is a direct reflection of the difference in the integrated kinetic power of
accretion shocks between these three models, as demonstrated in PF06.

\begin{figure}
  \centering
  \includegraphics[width=0.9\textwidth]{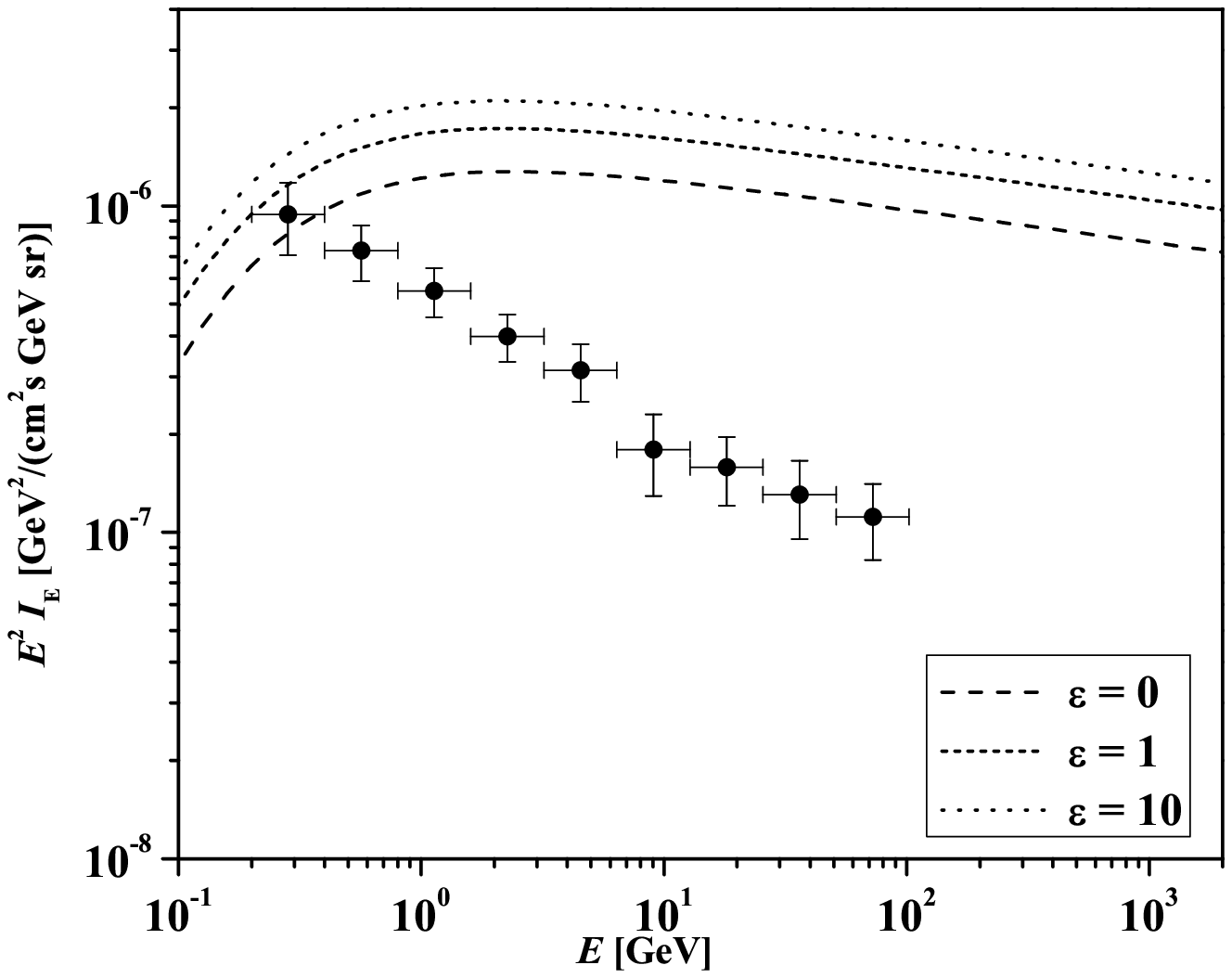}
  \caption{\label{fig:3}
This plot shows the sensitivity of our model on the adopted initial gas mass fraction parameter $\epsilon$. For the purpose of
demonstration, we plot the SFCR gamma-ray emission with spectral index $\alpha_\gamma=2.1$,
based on Model 1 of PF06, and derived adopting different initial gas mass fraction values, $\epsilon=0,1,10$. The top most curve
is approximately a factor of two higher than our fiducial case plotted in Figure \ref{fig:2}. For all $\epsilon > 10$, all curves converge
and are overlapping with the $\epsilon=10$ curve.}
\end{figure}

In Figure \ref{fig:3}, we demonstrate the sensitivity of our model with respect to the adopted initial gas mass content of a cluster
represented by the parameter $\epsilon$. We see that for the two most extreme cases, our fiducial case, $\epsilon=0$, and our limiting
case, $\epsilon=10$ (note that for $\epsilon>10$ the curves converge), the resulting curves differ
by a factor of $\sim 2$. To be as conservative as possible, we thus keep $\epsilon=0$ as our fiducial value.

\begin{figure}\centering
{\label{fig:4a}
\includegraphics[width=0.55\textwidth]{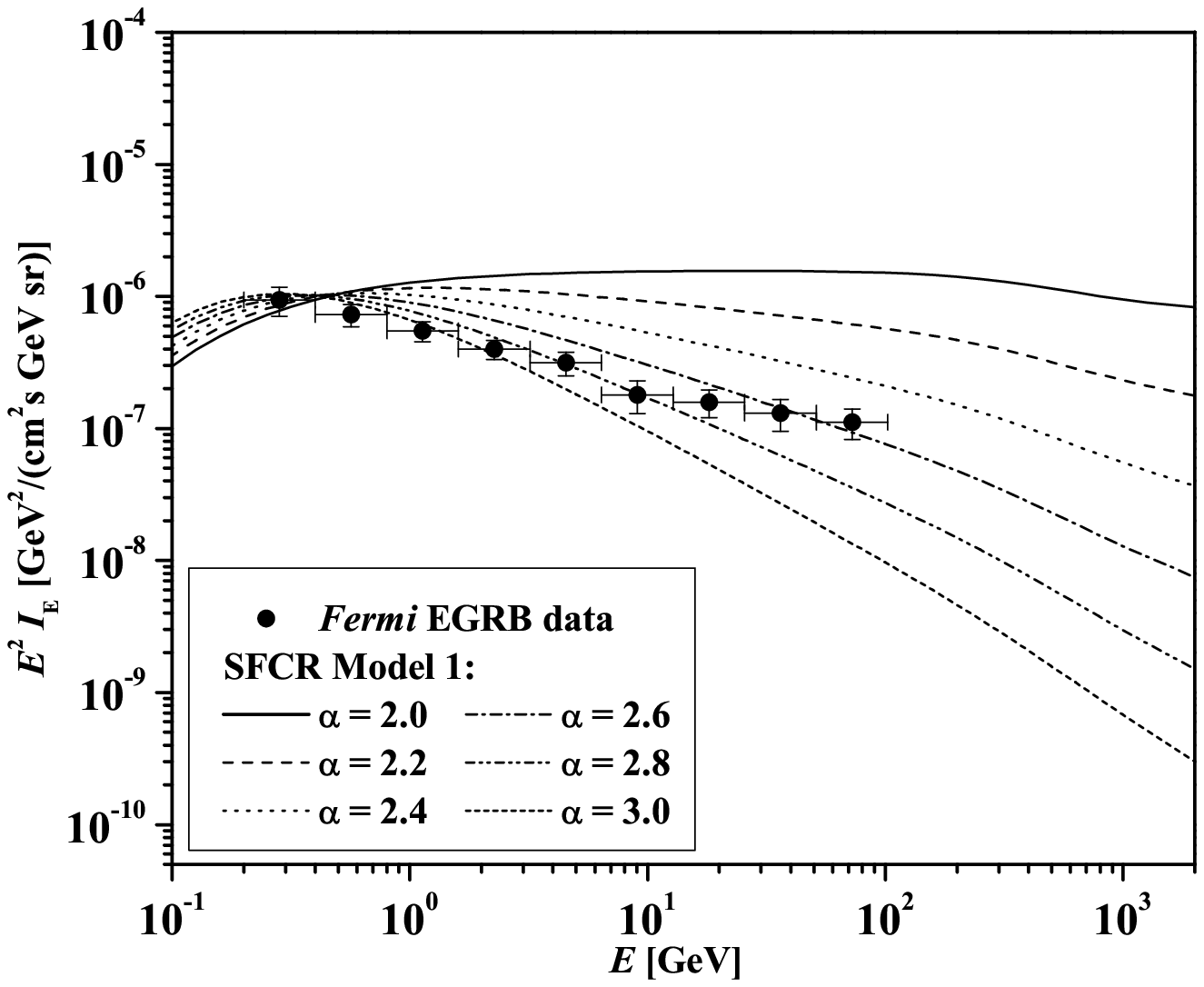}}
{\label{fig:4b}
\includegraphics[width=0.55\textwidth]{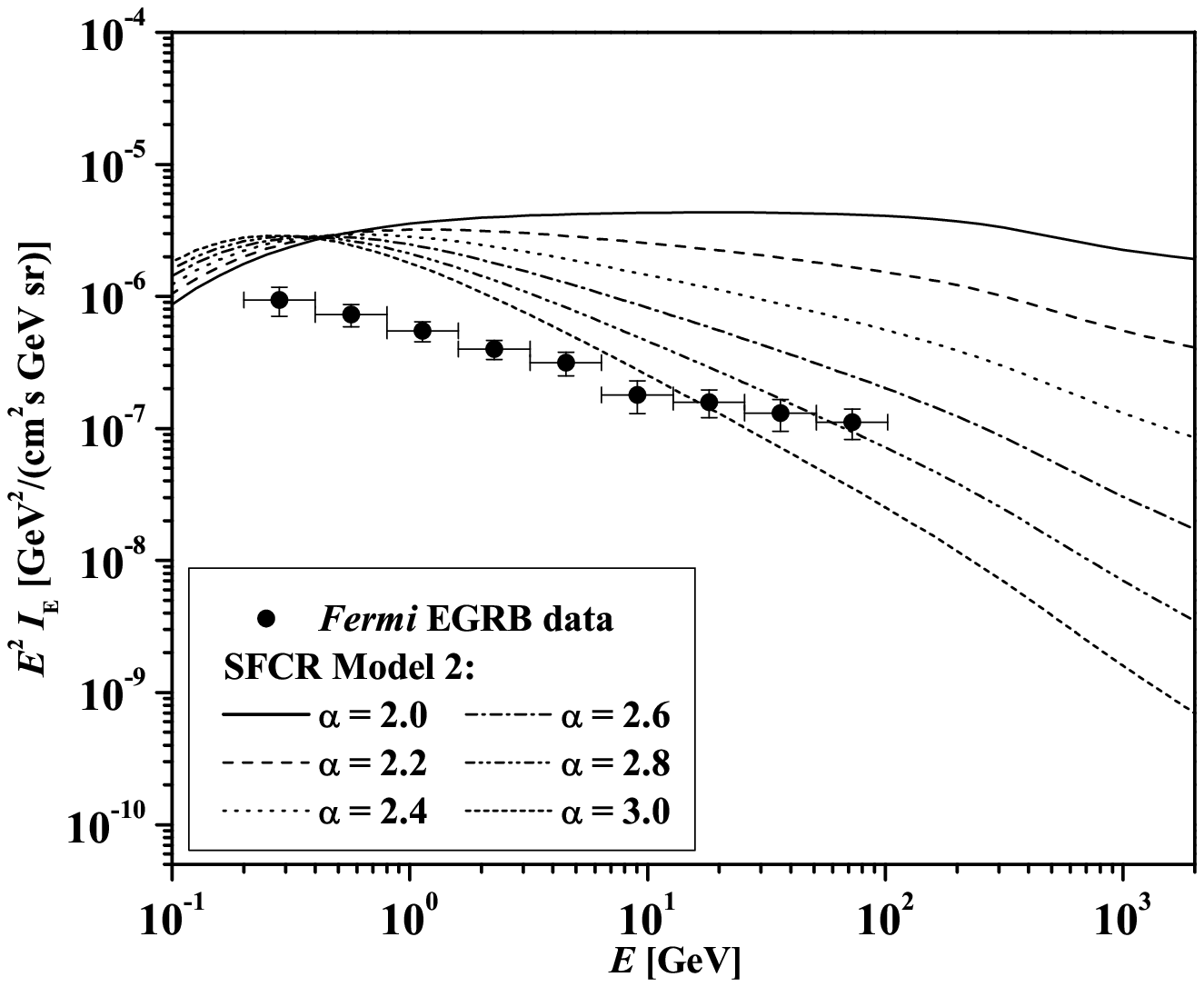}}
{\label{fig:4c}
\includegraphics[width=0.55\textwidth]{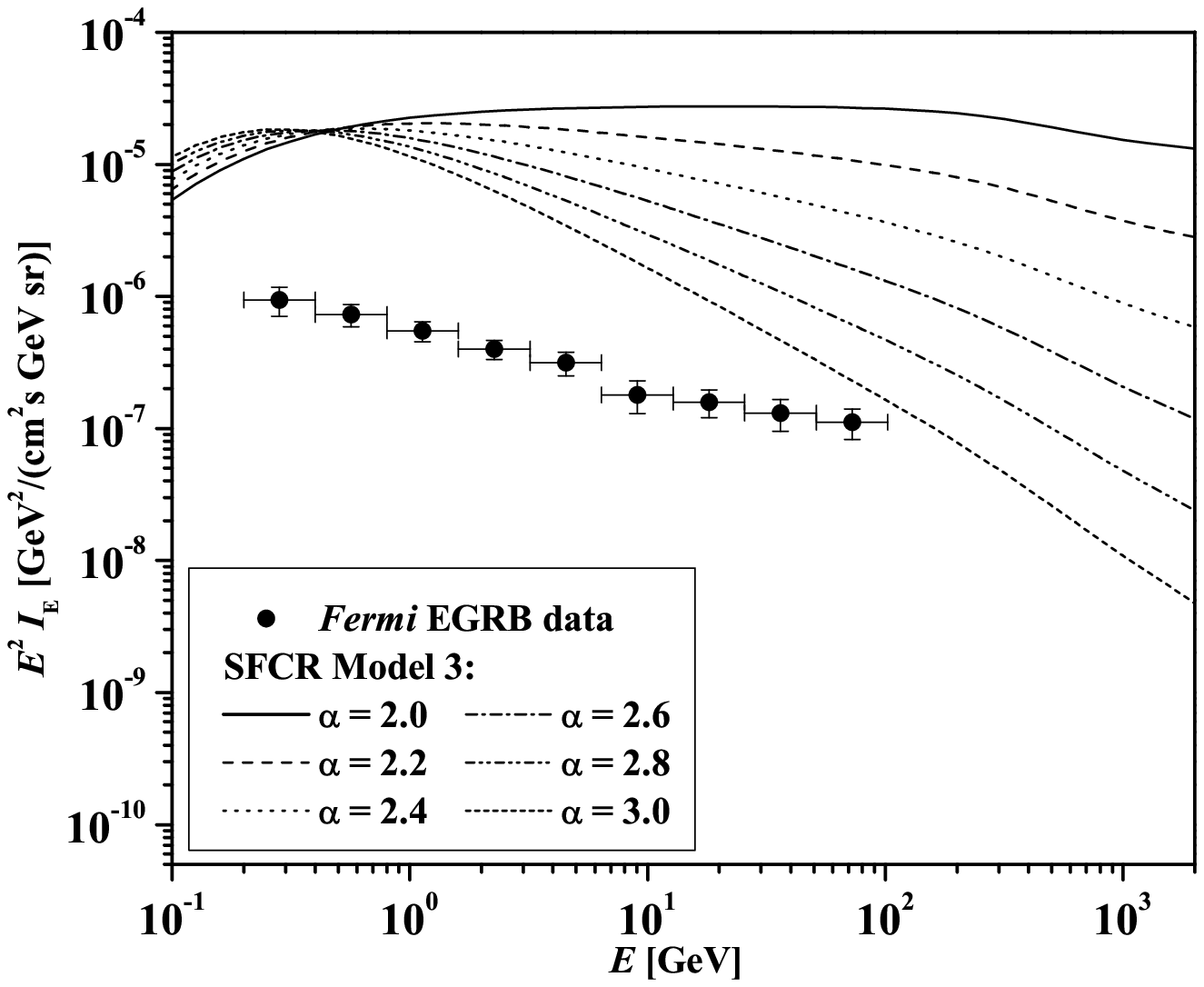}}
\centering
\caption{\label{fig:4}
Contribution of  structure-formation cosmic rays to the
EGRB (data points) observed by \emph{Fermi} \citep{ABD10} for different choices of the cosmic-ray spectral
index ranging from $\alpha_\gamma=2$ to $\alpha_\gamma=3$, for
 Model 1 (top panel), Model 2 (middle panel)  and Model 3 (bottom panel). The Coma cluster was again taken as the
 normalizing object with $\epsilon=0$.}
\end{figure}

In Figure \ref{fig:4}, we show how our results change with a different choice of the spectral index. The
SFCR gamma-ray component with spectral indices ranging from 2--3 is plotted for different PF06 models: Model 1, top pannel;
Model 2, middle pannel; and Model 3, bottom panel.

Our results for the fiducial case of strong shocks show that the observed EGRB is best matched with the SFCR component where
source evolution is based on the most simple model with no environmental effects taken into account, Model 1 of PF06, while
other models overshoot the observed data.
However, given that we have normalized our models to the upper limit of the Coma cluster emission, this is the
most generous estimate of the SFCR contribution to the EGRB. Once the cluster(s) has been detected, or if a detailed cluster
emission model is implemented, our upper limits would become predictions, but could also accommodate Models 2 and 3 with
softer emission.

\section{DISCUSSION AND CONCLUSION}
\label{sec:disc}

We have constructed a model of the collective gamma-ray emission arising from the large scale accretion
shocks around virialized structures and estimated its contribution to the extragalactic gamma-ray background
for various parameter values.
Assuming that accretion shocks give rise to a new population of cosmic rays, the SFCRs, this would inevitably
result in a gamma-ray flux which would contribute to the observed (but still unexplained!) EGRB at some level.
Given that SFCRs are still a hypothetical population, with no direct observational evidence and no known source evolution, so far only very
weak upper limits have been placed to their contribution to the EGRB.

Analytically modeling cluster formation and evolution, and including different sources of particle acceleration,
\citet{CB98} have estimated the cluster contribution to the gamma-ray background to be $ \la 1 \%$ of the EGRB measured by \emph{EGRET} \citep{ST00}, while numerical
models estimate it to be of the order of $ \la 10 \%$ \citep{MI02}. However, the latest measurements by \emph{Fermi} show the EGRB intensity above 100 MeV to be about a factor of 1.4 lower \citep{ABD10} than previously estimated, thus making the SFCR contribution to the gamma-ray background correspondingly larger.
This is especially important at higher energies where the \emph{Fermi} measured EGRB is almost an order of magnitude lower than inferred from \emph{EGRET}, e.g., making the \cite{ST00} estimate of the pionic SFCR component at the $\sim 40\%$ level of the EGRB.

In this work, we have implemented source evolution based on
the PF06 \citep{PF06} analytical model of cosmic accretion shocks around virialized structures within different environments, with
spectral index of the source gamma-ray emission as a free parameter.
The results of our analysis are, in some cases, consistent with previous estimates that this component is subdominant
to all other EGRB sources \citep{MI02,CB98,KU05}; however,
we also show that there are scenarios where accretion shocks on clusters can have an important contribution to
the EGRB.

To calibrate the resulting spectrum, we have used the unresolved source (point source) gamma-ray flux upper limits reported
by \emph{Fermi} for galaxy clusters
\citep{AC10}, specifically the Coma cluster; thus, our results presented in this work are the upper limits.
Moreover, by using these limits, our results are free of any assumption about the profile of the intracluster medium.
Taking into account a spatial profile would give a stronger limit, but we leave that analysis for the follow up work.

Our results, for our fiducial case of strong accretions shocks normalized to the Coma cluster, are presented in Figure \ref{fig:2}. We see that the contribution of cosmic rays, arising from accretion shocks, to the observed
EGRB can be dominant, especially at energies $> 10\,\rm GeV$ compared to cosmic rays originating from star-forming galaxies.
Moreover, depending on the assumptions, our discussed models go above the observed EGRB limits, allowing sufficient room for recalibration once
clusters have been detected by gamma-ray observations, or when a detailed model of cluster emission is implemented.
Thus, a positive cluster detection would not only make our upper limit model of SFCR contribution to the EGRB into a prediction, but would also
serve, within our model, as a probe of the SFCR source evolution. For the specific case of the Coma cluster, such detection might
be within reach, given that
new predictions \citep{Brunetti12}, which are at the same time successful in explaining the Coma cluster radio halo, fall just below
the present \emph{Fermi} limits.
On the other hand, our limits to SFCR gamma-ray emission can also be used to discriminate between different
models \citep{Brunetti12,PP10} of hadronic emission from the Coma cluster.
Recently, \cite{Keshet12} have reported a detection with VERITAS Cherenkov array of the gamma-ray ring around the Coma cluster.
The reported signal is claimed to be synchrotron and inverse-Compton emission from relativistic electrons accelerated in large-scale shocks.
Since the hadronic gamma-ray emission is thought to be subdominant in the reported signal, we cannot directly use this to
calibrate our model; however, if confirmed, such detection would be important for constraining the population of
SFCRs around clusters.

In terms of model dependence, our predicted limit is slightly sensitive to the choice of the virialization redshift of the accretor. Normalizing to a galaxy
cluster with a larger virialization redshift would result in a slightly higher collective gamma-ray emission and contribution to the EGRB.
On the other hand, the choice of a environment model for the cosmic accretion rate $\dot{\rho}_\mrm{sf}$ from PF06 can change
the results by about one order of magnitude. The impact of a different choice of the environment in the source evolution model is best evident at the low-energy end of
Figure \ref{fig:4}. Using the simplest model (Model 1) presented in PF06, which was based
only on the Press--Schechter distribution, where all objects accrete baryons of the uniform density and temperature, the resulting
spectrum is lower than in the case of a more realistic model with density and temperature variations around the accretor.
Even though it involves the most simplistic assumptions, we find that the choice of Model 1 for our source evolution, results
 in the SFCR gamma-ray component that best fits the observed EGRB.
Our results are also very sensitive to the choice of the spectral index of gamma-ray emission of a typical source, which
is best evident at the high energy end of Figure \ref{fig:4}.
This also results in variations by about an order of magnitude within a chosen environment model.
When more data at the high energy end becomes available, this will be important in determining the typical
environment and spectral index of a typical SFCR source.

Finally, our model is also strongly dependent on the choice of the galaxy cluster to which we normalize, which
can change the resulting spectrum up to two orders of magnitude, depending on the cluster size, distance, mass,
and its reported flux upper limit. Given the setup of our model, the normalizing cluster should be representative of the cosmic mean with respect to the cluster mass.
On the other hand, given that no cluster was actually detected and that only upper limits are available, ideally, the normalizing cluster should also be the one with its limit closest to its pionic gamma-ray emission. We have considered several clusters as candidates for the normalizing cluster, and eventually chose the Coma cluster. Though Coma is a rich cluster, it has intermediate values of cluster size, mass, and flux upper limit, where its predicted pionic emission is at the level of its \emph{Fermi} limit \citep{Brunetti12}. Other considered clusters were not as suitable as Coma for various reasons. For instance, Perseus is a cluster slightly smaller than Coma but its predicted flux upper limit is large because \emph{Fermi} sensitivity for this cluster is affected by foreground emission from the Galactic plane and other bright gamma-ray sources \citep{AC10}. Another considered candidate was the NGC5813 cluster, which is smaller than Coma with mass $M_\mathrm{500}=4.3\times10^{13}\,M_{\bigodot}$ \citep{CH07} and closer to the cosmic average. However, this cluster has a weaker \emph{Fermi} limit, and thus, the resulting upper limits to the SFCRs contribution to the EGRB  would be weaker by a factor of $\approx 1.5$ than in the case of Coma.
On the other hand, Fornax is a cluster with mass $M_{500}=1.24\times10^{14}\,M_{\bigodot}$\citep{CH07}, which is around eight times smaller than $M_{500}$ for Coma; however it is expected to be dominated by the dark matter annihilation signal \citep{PPB11} and thus would not be a good choice for normalization of our model.

Even though our new constraints on the SFCR contribution to the EGRB are quite model-dependent and represent limits, the results
indicate that there is still possible importance of this gamma-ray component which could have multiple implications.
For instance, a concern was raised that SFCRs
could potentially produce important quantities of lithium isotopes which would increase the severity of the
lithium problem \citep{PF07,SI02}. Since we have shown that the SFCR contribution to the EGRB can be important, corresponding
lithium production \citep{FP05} could also be relevant.
Similarly, neutrino fluxes accompanying this cosmic-ray population could
also make an important contribution to the neutrino background arising from other sources like dark matter annihilations
\citep{CB98,MB12}. Therefore, it is not only important to detect or build a careful model of accretion shock cluster
emission, but to also let the sources evolve within a carefully treated environment, because this can greatly affect the resulting
gamma-ray emission.

\acknowledgments

We are grateful to Vasiliki Pavlidou and Brian Fields for insightful discussions and comments and valuable help with data.
We are also thankful to the anonymous Referee whose comments helped improve this paper.
The work of AD is supported by the
Ministry of Science of the Republic of Serbia under project number 176005 and the work of TP is supported in
part by the Ministry of Science of the Republic of Serbia under project numbers 171002 and 176005.

\newpage

\center{\bf ERRATUM: "DIFFUSE PIONIC GAMMA-RAY EMISSION FROM LARGE-SCALE STRUCTURES IN THE \emph{FERMI} ERA" (2014, ApJ, 782, 109)}

\author{ A. Dobard\v zi\'c}
\center{\it Department of Astronomy, Faculty of Mathematics, University of Belgrade, Studentski trg 16, 11000 Belgrade, Serbia}
\center{\verb"aleksandra@matf.bg.ac.rs"}

\and

\author{T. Prodanovi\'c}
\center{\it Department of Physics, University of Novi Sad, Trg Dositeja Obradovi\'ca 4, 21000 Novi Sad, Serbia}
\center{\verb"prodanvc@df.uns.ac.rs"}

\justifying

The published version of this article contained a computational error in calculation of the mass accretion rate $J_0$ (mass current crossing
the shock surface in units of $M_\odot \rm yr^{-1}$) at redshift $z_0$ of the galactic cluster to which we normalize our models
of structure formation cosmic rays. The mass accretion rate $J_0$ was calculated using \cite{PAFI06}:

\begin{equation}
J_0(z_0)=4\pi r_\mathrm{v}^2(m)\Omega_\mathrm{b}\rho_\mathrm{c,0}(1+z_0)^3(1+\delta_\mathrm{s}){\cal M}c_\mathrm{s,1}\,,
\end{equation}
where $\Omega_\mathrm{b}=0.04$ is the baryonic matter energy density parameter, $\rho_\mathrm{c,0}$ is the critical
density at the present epoch, $c_\mathrm{s,1}$ is the adiabatic sound speed of the pre-shocked material, $\delta_\mathrm{s}$
is the overdensity in which the accretor is located, $r_\mathrm{v}$ is the virial radius of the accretor. Mach number of the
accretion shock is ${\cal M}$. We chose to normalize to Coma cluster, so the correct value of the accretion rate for this cluster
is $J_0=417.86\,M_\odot\mathrm{yr}^{-1}$. This value is an order of magnitude lower than the value used in the published
version of the paper. This does not change our main conclusion that structure formation cosmic rays can make an important
contribution to the extragalactic gamma-ray background, but it does change the parameter values of the best-fit model.
All of our models now give an order of magnitude higher gamma-ray flux, which now puts a tighter constraints on the ones
that are allowed by the date.

The observed extragalactic gamma-ray background is, as in the published version, best matched by the structure formation
cosmic-ray component where source evolution is based on the most simple
model with no environmental effects taken into account, Model 1 of \cite{PAFI06}, while other models overshoot the observed
data. In the Results section of the published version we have adopted as our default case the spectrum where the initial gas mass
parameter was taken to be $\epsilon = 0$,
while the adopted spectral index was that of the strong shocks $\alpha_\gamma = 2.1$. Since our curves are now higher,
even our simplest model is above the
extragalactic gamma-ray background observed by the \emph{Fermi} for strong shocks
spectral indices $\alpha_\gamma\approx 2$. Our most probable scenario is now for the spectral index $2.7$ (Model 1).
This is close to \cite{BBR12} where they found best-fit of the Coma spectrum to be derived using spectral index $\approx2.6$.
Although even in this case of the softer spectra, the model does overshoot the data at some energy ranges, we
remind that our normalization was based on the Coma gamma-ray upper limit as reported by \emph{Fermi}, and
thus leaves room for downward correction once the detection is made or when a specific emission model is used.

We plot Figure 1 with the same parameters, just like in the published version of the paper, but with now corrected value of $J_0$
(top panels). Here we also add the same plots of our best-fit scenario with spectral index $2.7$ (bottom panels). Figure 1 shows
contributions of structure formation cosmic rays, normal galaxies and blazars shown separately (left), as well as their summarizes
contribution (right). We also correct Figure 2 from our published paper using our
best-fit spectral index $2.7$ instead of $2.1$ which was used before. Figure 3 is also corrected and plotted using the same
parameters like in the published version.

Our models still show that structure formation cosmic rays can make a substantial contribution to the extragalactic gamma-ray
background, but the data are now more constraining to our models. Resulting contributions are now higher than those derived in
some of the papers mentioned in the Discussion and Conclusion section of our published article \citep{MI02,CB98,KU05}.
\\
\\

\begin{figure}[h!]\centering
{\label{fig:1}
            \includegraphics[width=0.9\textwidth]{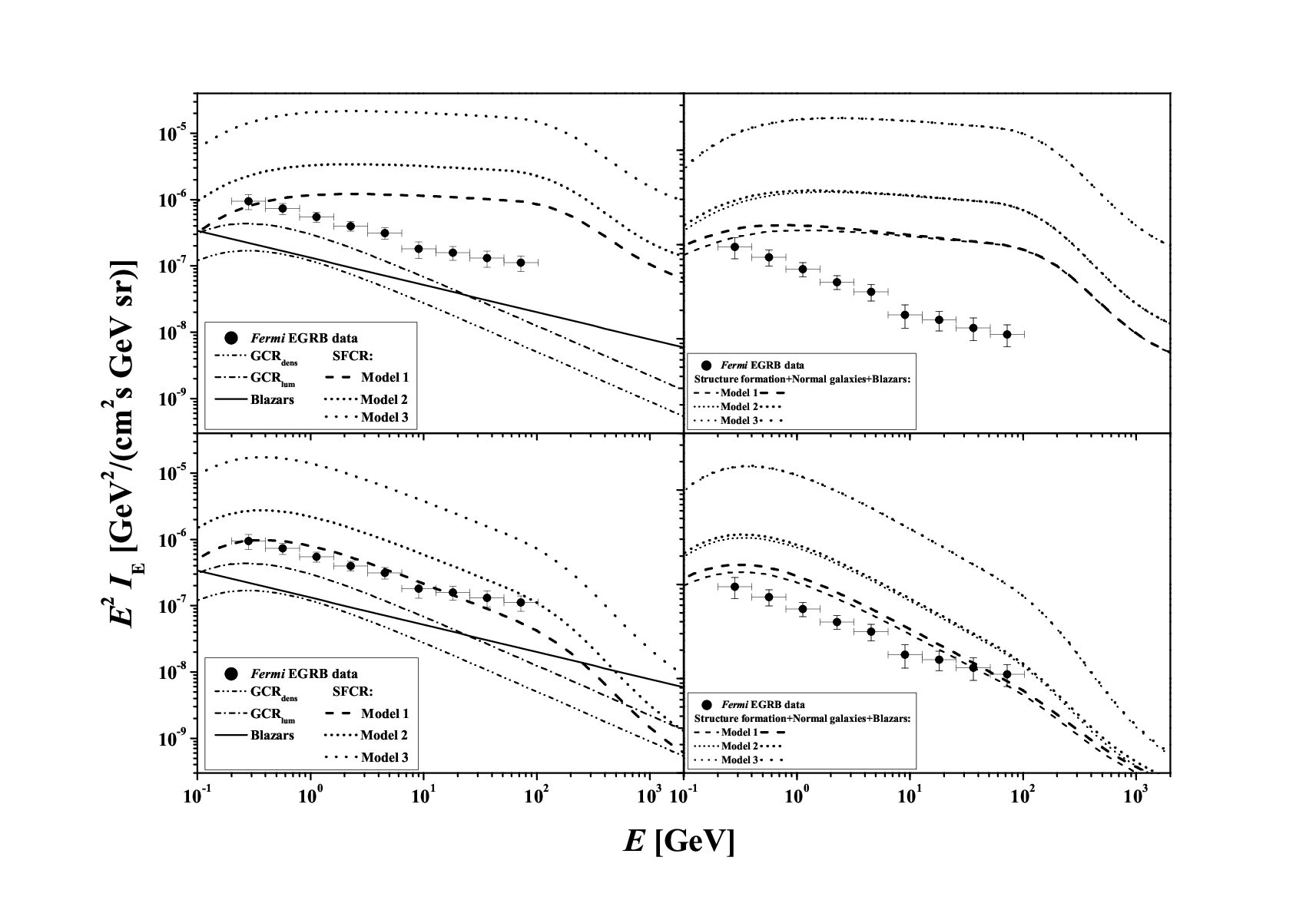}}
\centering
    \caption{\label{fig:1}
Contribution of different components to the
extragalactic gamma-ray background (data points) observed by \emph{Fermi} \citep{ABD10}.
{\it Left:} all components shown separately -- blazars (solid line), normal star-forming galaxies based on two limiting cases given in
Fields et al. (2010; red dash dotted line represents luminosity evolution and blue dash dot dotted line represents density evolution),
and structure formation cosmic-ray contribution calculated as in the published article, but with the corrected value for $J_0$,
normalized to the Coma cluster gamma-ray flux limit, with initial gas mass parameter $\epsilon=0$, for three different source
models derived in Pavlidou \& Fields (2006; long dashed, Model 1; short dashed, Model 2; dotted line, Model 3). Top panel shows
structure formation cosmic-ray spectra derived using spectral index $\alpha_\gamma=2.1$ and bottom panel $\alpha_\gamma=2.7$.
{\it Right:} the combined contribution of all components where different curves reflect different
normal galaxy emission models (thick red curves, luminosity evolution; thin blue curves, density evolution)
and different structure formation cosmic-ray emission models
(three different line types correspond to the same models as on the top panel). Top panel shows structure formation cosmic-ray
spectra derived using spectral index $\alpha_\gamma=2.1$ and bottom panel $\alpha_\gamma=2.7$.
\newline
(A color version of this figure is available in the online journal.)}
\end{figure}

\begin{figure}[h!]\centering
{\label{fig:2}
            \includegraphics[width=0.55\textwidth]{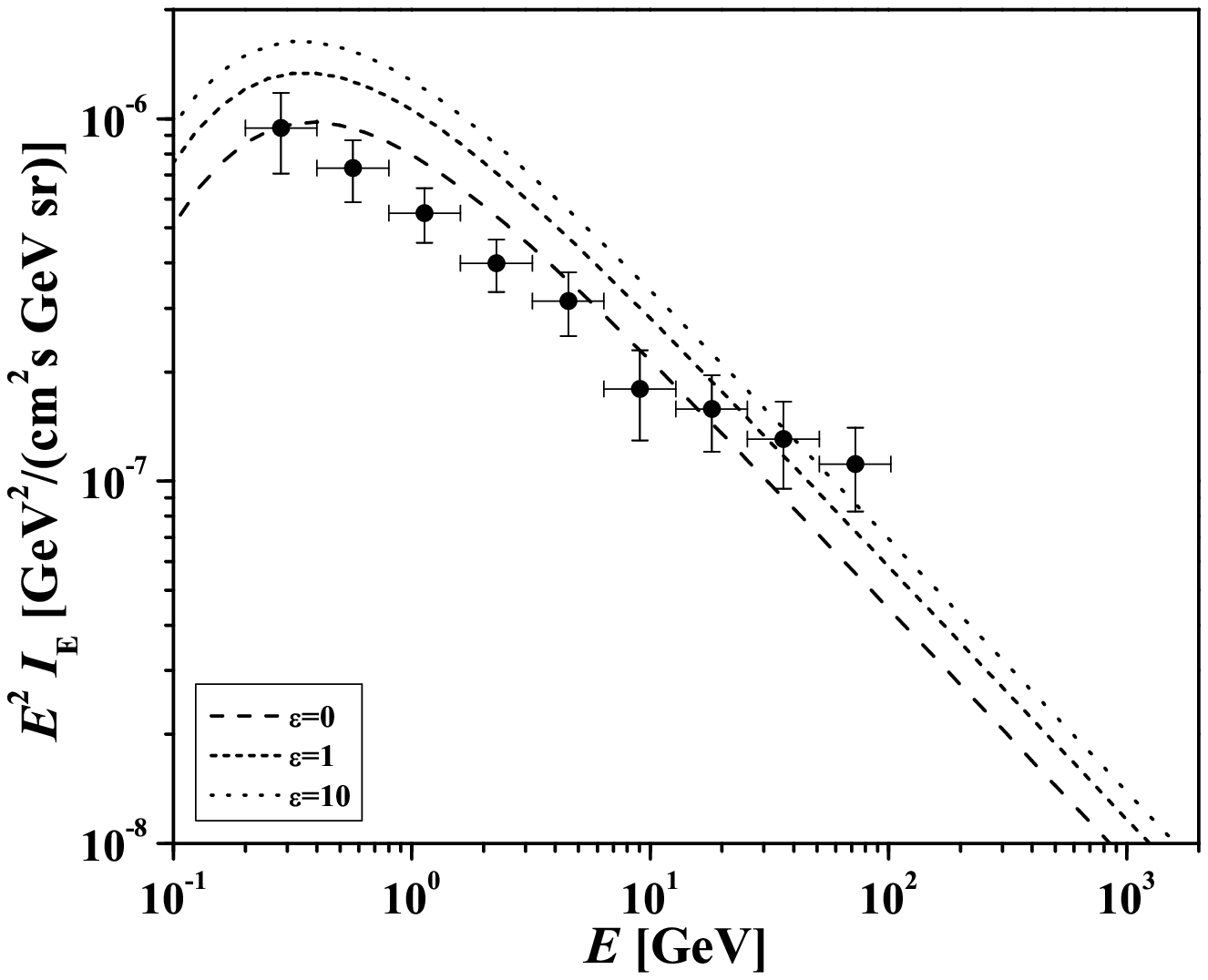}}
\centering
\caption{\label{fig:2}
This plot shows the sensitivity of our model on the adopted initial gas mass fraction parameter $\epsilon$. For the purpose of
demonstration, we plot the structure formation cosmic-ray gamma-ray emission for out best-fit model with spectral index
$\alpha_\gamma=2.7$, based on Model 1 from \cite{PAFI06}, and derived adopting different initial gas mass fraction values,
$\epsilon=0,1,10$. The top most curve is approximately a factor of two higher than our best-fit case plotted in Figure \ref{fig:1}.
For all $\epsilon > 10$, all curves converge and are overlapping with the $\epsilon=10$ curve.}
\end{figure}

\begin{figure}[h!]\centering
{\label{fig:3}
\includegraphics[width=0.55\textwidth]{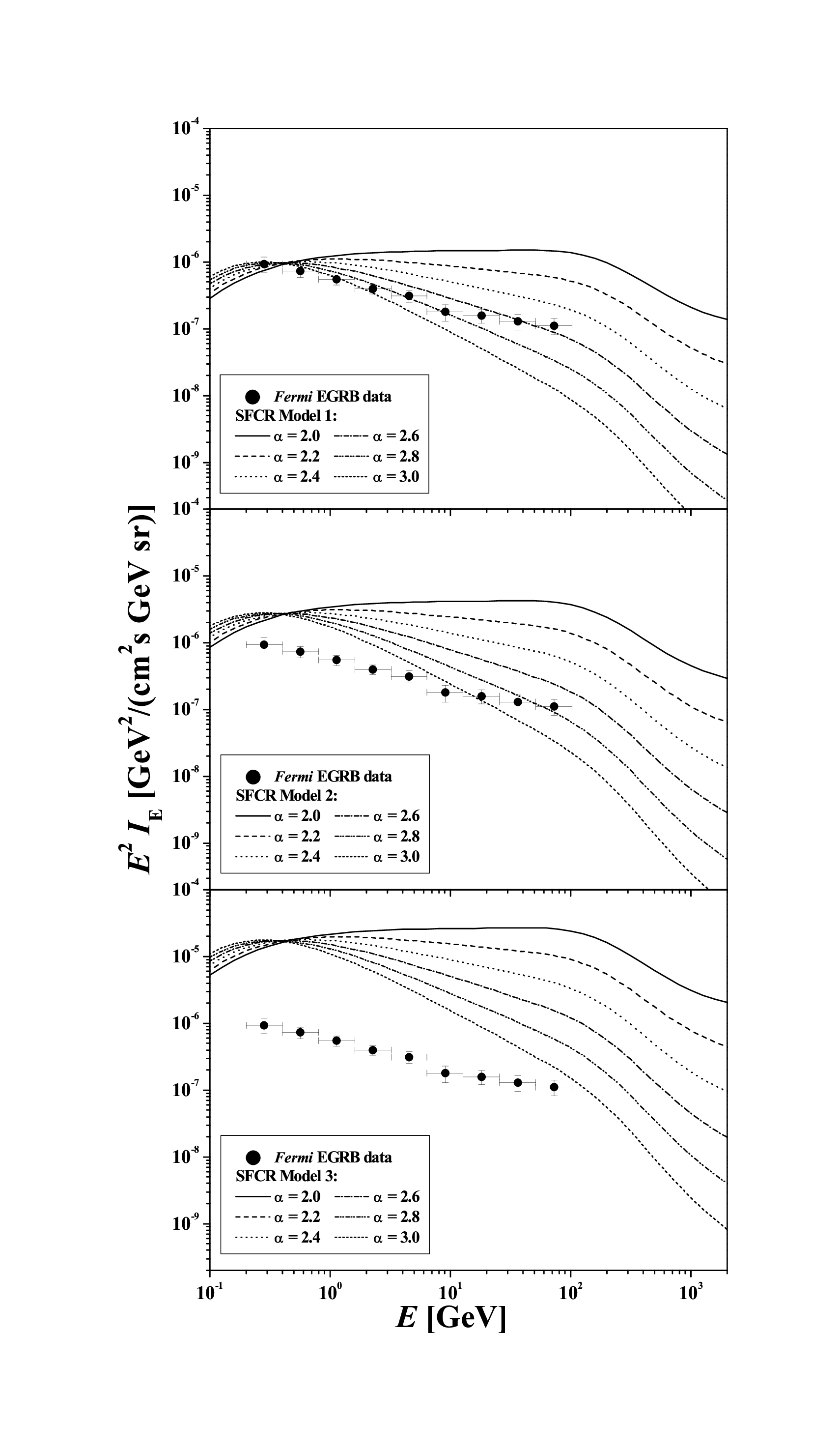}}
\centering
\caption{\label{fig:3}
Contribution of  structure formation cosmic rays to the
extragalactic gamma-ray background (data points) observed by \emph{Fermi} \citep{ABD10} for different choices of the cosmic-ray spectral
index ranging from $\alpha_\gamma=2$ to $\alpha_\gamma=3$, for
 Model 1 (top panel), Model 2 (middle panel)  and Model 3 (bottom panel). The Coma cluster was again taken as the
 normalizing object with $\epsilon=0$.}
\end{figure}

\end{document}